\documentclass{aa}  
\usepackage{graphicx}
\usepackage{natbib}
\usepackage{txfonts}
\newcommand{\degrees}[1]{\ensuremath{#1^\circ}}

\begin{document}
   \title{The main- interpulse interaction of PSR B1702--19}


   \author{P. Weltevrede
          \inst{1}
          \and
          G.~A.~E. Wright\inst{2}
	  \and
	  B.~W. Stappers\inst{3,1}
          }

   \offprints{P. Weltevrede}

   \institute{Astronomical Institute ``Anton Pannekoek'',
              University of Amsterdam,
              Kruislaan 403, 1098 SJ Amsterdam, The Netherlands\\
              \email{wltvrede@science.uva.nl}
         \and
             Astronomy Centre, University of Sussex, Falmer, BN1 9QJ, UK\\
             \email{G.Wright@sussex.ac.uk}
	 \and
	     Stichting ASTRON, Postbus 2, 7990 AA Dwingeloo, The Netherlands\\
             \email{stappers@astron.nl}
             }

   \date{Received ...; accepted ...}

 
  \abstract
   {}
   { This paper reports on single-pulse radio observations of PSR B1702--19 
and their implications for pulsar emission theories. }
   { These observations were made with the Westerbork Synthesis Radio Telescope 
at 1380 and 328 MHz.  The PA-swing is used to constrain possible geometries of
the pulsar and the single-pulse data is analysed for subpulse modulation 
correlations between the main pulse and interpulse.}
   { We confirm earlier conclusions that the dipole axis of this pulsar is
almost perpendicular to its rotation axis, and report that both its
main pulse and interpulse are modulated with a periodicity around $10.4$
times the pulsar's rotation. Allowing for the half-period delay
between main pulse and interpulse the modulation is found to be
precisely in phase. Despite small secular variations in the
periodicity, the phase-locking continues over all timescales ranging
up to several years. }
   { The precision of the phase locking is difficult for current
emission theories to explain if the main pulse and interpulse
originate from opposing magnetic poles. We therefore also explore the possibility of a
bidirectional model, in which all the modulated emission comes from one
pole, but is seen from two sides and slightly displaced by aberration
and time-delay. In this model the unmodulated emission is
directed to us from the opposite pole, requiring the emission of the
main pulse to originate from two different poles. This is difficult to
reconcile with the observed smooth PA-swing. 
Whichever
model turns out to be correct, the answer will have important
implications for emission theories. }


\maketitle

\section{Introduction}
The radio pulsar B1702--19 was discovered by \cite{mlt+78} in the
second Molonglo Survey. With a period of 0.3  seconds, a
characteristic age of 1.1 Myr and inferred surface magnetic field
$1.1\times10^{12}$ Gauss, the pulsar was in no way unusual. However,
some ten years later it was found to have an interpulse located some
\degrees{180} from the main pulse (\citealt{blh+88}), an unusual feature
among pulsars, and one which has the potential for testing
magnetosphere and emission models.

Radio pulsars have narrow emission beams, so that an interpulse
(henceforth IP) located close to \degrees{180} from the main pulse
(henceforth MP) might naturally be interpreted as a view of a second
magnetic pole. Previous studies of PSR B1702--19 have tended to
support this view. For example, according to \cite{blh+88}, there is
little evidence for the MP-IP separation (given as
$\degrees{181}\pm\degrees{1}$) to evolve with frequency. Further
support is given by the analyses of the shape of the polarisation
position angle swing in \cite{lm88} and \cite{kuz89}, which is more
recently confirmed by \cite{vdhm97}. However, geometric models exist
(e.g. \citealt{gil85}, \citealt{dzg05}) in which even a separation of
this magnitude can arise from a single pole.

The structure of the MP is essentially double-peaked at all
frequencies, the leading peak being more than \degrees{180} from the
following IP and the trailing peak less than \degrees{180}. The
midpoint of the MP is therefore more-or-less exactly at the
\degrees{180} point. There is a weak shoulder on the trailing edge, so
that the MP profile has been classified as triple by e.g.
\cite{ran90}.
\cite{kj04}, \cite{gl98} and \cite{blh+88} also note a high degree
of circular polarisation, especially at the MP. The IP itself seems to
be single-peaked and almost 100\% linearly polarised
(e.g. \citealt{kj04}).

In a recent survey of pulsar modulation at an observing wavelength of
21 cm, \cite{wes06} found that both the MP and IP of PSR B1702--19
have apparently identical periodic intensity modulations with a
periodicity ($P_3$) of about $11P$, where $P$ is the rotation period
of the neutron star. This exciting result immediately suggested a
method for examining interactions between the MP and IP and thereby
constraining emission theories, models for subpulse drift, and
magnetospheric structures.

Evidence of communication between the poles already exists in PSRs
B1822--09 and B1055--52\footnote{It is arguable that PSR B0950+08 also
displays inter-pole communication since emission at its MP is
correlated with the following IP (\citealt{hc81}). As in PSR
B1055--52, the IP precedes the MP by $\sim\degrees{150}$, but is
connected to it by a clear bridge of emission. It is still a matter of
debate whether this pulsar is a perpendicular rotator (\citealt{ew01})
or close to alignment (e.g. \citealt{hx97}).}. The
former is strikingly similar to PSR B1702--09, having a double-peaked
MP, a weak IP
\degrees{180} away and  a modulation with the same periodicity at both
longitudes (\citealt{fmw81}, \citealt{fw82},
\citealt{gjk+94}). However, PSR B1822--09's most remarkable feature is the
anticorrelation between the leading component of the MP and the IP, so
that the pulsar frequently switches between two modes, one displaying
both components of the MP, the other only the second MP component and
IP emission. An apparent geometric resolution of this phenomenon has
been published (\citealt{dzg05}), but as yet lacks a physical
explanation. In PSR B1822--09 the weakness of the IP has hitherto
prevented any precise study of potentially revealing phase differences
between the MP and IP modulations.

PSR B1055--52, like both PSRs B1702--19 and B1822--09, has a MP with a
strong first component and an overlapping double second component
(e.g. \citealt{mhak76}), and again the first component appears to
undergo mode-changes (\citealt{big90a}).  But the interaction with the
IP is less clear-cut: correlations of roughly one phase delay exist
between the intensities of the IP and the following MP, but no
comparable periodicities could be established in the modulations at
either poles.  Nonetheless, inter-pole communication seems to be
established in both these pulsars and it represents a challenge to
contemporary pulsar theory.

In this paper we set out to investigate the single-pulse behaviour of
PSR B1702--19 and to establish the nature of the interactions between
its MP and IP. We also try to assess the significant theoretical
constraints implied by our results. In Sect. \ref{SctObs1702} our
observations are listed. In Sect. \ref{SctProf} the profiles of both
MP and IP at two frequencies are presented and a possible mode-change
identified. In Sect. \ref{SctSinglePulses} we give a detailed
analysis of the single pulse modulations in the MP and IP, and
demonstrate their correlations.  In Sect. \ref{SctPA} we use
polarisation data to establish the likely viewing geometry of the
observer and PSR B1702--19. Finally in Sect. \ref{SctDisc1702} we
suggest possible geometric configurations for the MP and IP beams and
discuss their theoretical implications, before summarising our
principal results in Sect. \ref{SctSummary}.

\section{\label{SctObs1702}Observations}

\begin{table}
\caption{\label{UsedObservationsTable1702}The details of the observations
used in this paper. Here REF is the reference key used in this paper,
MJD is the modified Julian date of the observation, $\nu$ is the
central frequency of the observation, $\Delta\nu$ the bandwidth,
$\tau_\mathrm{samp}$ the sampling time, $N$ the number of recorded
pulses and the last column states if the observation contains
polarization information.}
\begin{center}
\begin{tabular}{c|cr@{}lr@{}lr@{.}lcc}
\hline
\hline
REF & MJD & \multicolumn{2}{c}{$\nu$} & \multicolumn{2}{c}{$\Delta\nu$} & \multicolumn{2}{c}{$\tau_\mathrm{samp}$} & $N$ & Pol.\\
&  & \multicolumn{2}{c}{(MHz)} & \multicolumn{2}{c}{(MHz)} & \multicolumn{2}{c}{(ms)} &  &\\
\hline
2006P & 53987 & \hspace*{0.8mm}328& & \hspace*{2mm}10&  & 0&2048 & 11885 & N\\
2003L & 53003 & \hspace*{0.8mm}1380& & \hspace*{2mm}80&  & 0&4096 & 11071 & Y\\
2005L1 & 53684 & \hspace*{0.8mm}1380& & \hspace*{2mm}80&  & 0&2048 & 18918 & N\\
2005L2 & 53684 & \hspace*{0.8mm}1380& & \hspace*{2mm}80&  & 0&2048 & 23935 & N\\
\hline
\end{tabular}
\end{center}
\end{table}

PSR B1702--19 was observed several times over three and a half years
with the Westerbork Synthesis Radio Telescope (WSRT) at an observing
wavelength of both 21 and 92-cm. The important parameters of the
observations that are used in this paper can be found in Table
\ref{UsedObservationsTable1702} and each observation is given a
reference key. The 2003L and the 2006P observations were part of the
subpulse modulation survey done with the WSRT
(\citealt{wes06,wse07}). The other observations were carried out at
identical radio frequencies and were reduced using the same procedure
as for the survey, so for a more detailed description of the
data-reduction we refer to those papers.

The two 2005 observations were made directly after each other with
only a few minutes between them. To obtain a high signal-to-noise
($S/N$) profile, we summed the two integrated pulse profiles of the
two observations together. The profiles were aligned by correlating
the two profiles, which could be done with an accuracy of the sampling
time (0.2 ms). We will use the reference key 2005L for the summed
profiles of the 2005L1 and 2005L2 observation.

\section{\label{SctProf}The pulse profile of PSR B1702--19}
\subsection{The total intensity profile}

\begin{figure}
\begin{center}
\resizebox{0.99\hsize}{!}{\includegraphics[angle=0,trim=0 0 0 0,clip=true]{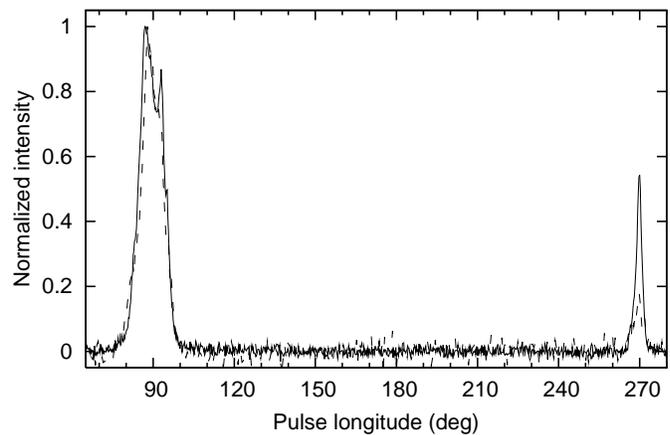}}
\end{center}
\caption{\label{profiles_two_freq}The intensity profiles of
the 1380 MHz 2005L observation (solid line) and the 328 MHz 2006P
observation (dashed line). The time-resolution of the latter is
reduced by a factor two. The peak of the IP is, in both cases, put on
\degrees{270} pulse longitude.  The two profiles are
normalized to their peak intensity.}
\end{figure}

The pulse profiles at 1380 and 352 MHz are shown in
Fig. \ref{profiles_two_freq}. The IP is relatively strong compared
with the MP at 1380 MHz, showing that the spectra of the MP and IP are
different. While the IP is single-peaked at both frequencies, the MP
is double peaked, especially at high frequencies. The width of the MP
does not evolve much with frequency, except that the separation
between the two peaks of the MP becomes smaller at low frequencies and
moreover they seem to merge.

The separation between the MP and IP is, within the noise level, the
same at both frequencies and at 1380 MHz the peak-to-peak separation
is $\degrees{183}\pm0.5$. This is consistent with the absence of a
frequency dependence of the MP-IP separation of $181\pm\degrees{1}$
reported by \cite{blh+88}. Interestingly, the peak of the IP appears
to be exactly \degrees{180} from the midpoint of the two peaks of the
MP. This would be in agreement with the two-pole scenario in which the
magnetic axes are separated by \degrees{180} and the magnetic axis
coincides with the peak of the IP and the other axis with the centre
of the emission cone that produces the MP. Also
\cite{vdhm97} report a centroid MP-IP separation of
$180.1\pm\degrees{0.4}$.

In the single wide-cone scenario (\citealt{ml77}) the MP
and IP are produced by the same emission cone. In that case it would
be coincidence that the component separation is very close to
\degrees{180}. The presence of a ``bridge'' between the two components
would be an argument in favor of a wide cone. However there is no sign
for weak emission between the two components. The detection limit on
the peak-flux ($3\sigma$) of any low-level bridge between the
components is 1\% of the MP intensity at 1380 MHz. There is no
indication for an anti-symmetry in the shape of the MP and IP, which
could have been an argument in favor of a wide-cone scenario.

Notice also that there is some indication of a third MP component at
high frequencies at pulse longitude $\sim$\degrees{95} in
Fig. \ref{profiles_two_freq}. This would be consistent with the
triple classification made by \cite{ran90}, also noted by
\cite{gl98}. Moreover, it will be shown in Sect. \ref{SctPolarization} 
and \ref{SctSinglePulses} that there is also evidence from the linear
polarization and the pulse modulation that there is indeed an addition
component at that pulse longitude.

It must be noted that there is some evidence that PSR B1702--19
occasionally changes its mode of emission, because the pulse profile
published by \cite{sgg+95} is significantly different from other
published profiles at similar frequencies.  It is striking that the MP
trailing component in their 1315 MHz profile is the strongest.
Unfortunately the Effelsberg data-set does not contain single pulses,
which makes it impossible to identify the possible mode-change.

\subsection{\label{SctPolarization}The polarization profile}

\begin{figure}
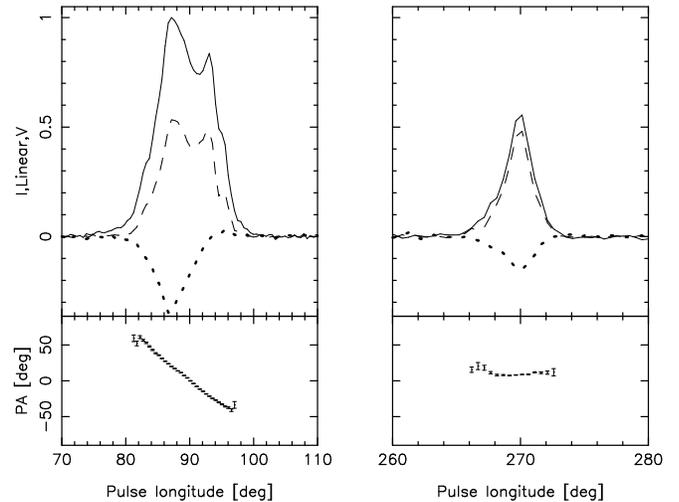

\begin{center}
\resizebox{!}{0.75\hsize}{\includegraphics[angle=0,trim=0 0 0 0,clip=true]{6957f2a.ps}}
\hspace*{0.05\hsize}
\resizebox{!}{0.75\hsize}{\includegraphics[angle=0,trim=0 0 0 0,clip=true]{6957f2b.ps}}
\caption{\label{polprofile}The left panel shows the MP of the 1380-MHz 2003L 
observation and the right panel the IP. The pulse profile in total
intensity (solid line), linear polarization (dashed line) and circular
polarization (dotted line) are normalized to the peak intensity of the
MP. The bottom panel shows the PA-swing. The pulse longitude of the
peak of the IP is set to \degrees{270} and at \degrees{90} the PA is
set to zero.}
\end{center}
\end{figure}

The polarization profiles of PSR B1702--19 at 1380 MHz are shown in
the top panels of Fig. \ref{polprofile}. As one can see, the degree of
linear polarization is quite high, especially in the IP. Also
\cite{kj04} find that the IP is almost 100\% linearly polarized at
1.41 GHz.  Notice that the third component of the MP is visible in the
linear polarization profile as an extra component. Also the degree of
circular polarization is found to be very high for this pulsar
(e.g. \citealt{gl98,blh+88}), something that is confirmed in the WSRT
data.

The emission of the MP of PSR B1702--19 is classified differently by
various authors as core and conal emission.  The emission of the MP is
classified by \cite{lm88} as cone-dominated. However, it is pointed
out by \cite{kj04} that the change in handedness of the circular
polarization in the MP suggests that the entire MP is a core
component. This in contrast to the classification made by
\cite{ran90}, who classified the MP as a core component with conal
outriders making it a triple profile, while the IP is classified as a
core single component.

\section{\label{SctSinglePulses}Pulse Modulation}
\subsection{The single pulse modulation}

\begin{figure}
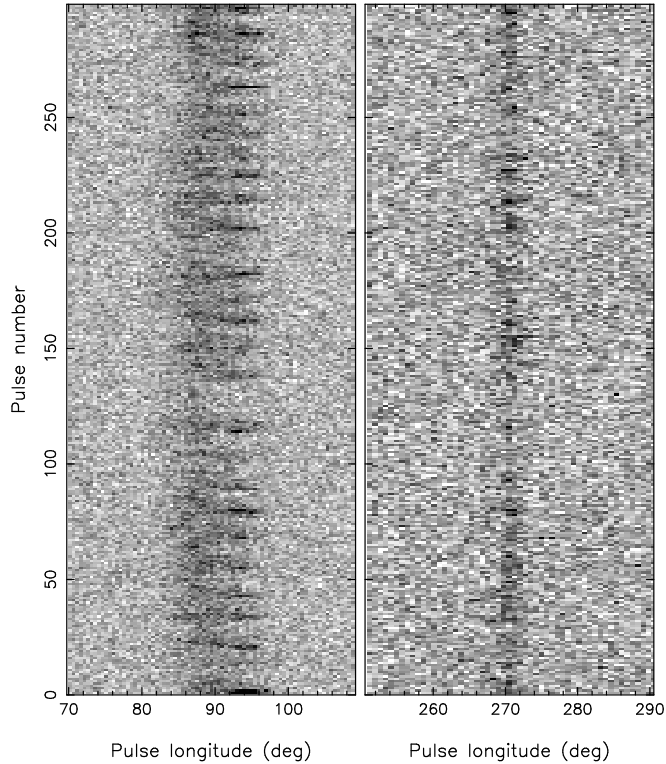

\begin{center}
\resizebox{!}{1.15\hsize}{\includegraphics[angle=0,trim=0 0 0 0,clip=true]{6957f3a.ps}}
\resizebox{!}{1.15\hsize}{\includegraphics[angle=0,trim=0 0 0 0,clip=true]{6957f3b.ps}}
\end{center}
\caption{\label{pulsestack}This is a pulse-stack of 300 successive
pulses from the 2005L1 observation. The left and right panel shows
respectively the longitude range around the MP and IP. The
time resolution of the left and right panels are reduced with a factor
two and three respectively.  Notice that the IP shows,
similar to the trailing half of the MP, a $P_3\simeq10P$ modulation.}
\end{figure}

It has been found by \cite{wes06} that the single pulses of both the
MP and IP show subpulse modulation and in Fig. \ref{pulsestack} a
bright piece of data is shown to illustrate this. The data in this
figure is represented as a pulse-stack, i.e. the successive pulses are
plotted on top of each other. The left and right panels show the
single pulses of respectively the MP and IP for the same rotation of
the star. The MP is clearly modulated with a periodicity of about
$10P$. The modulation is very strong in the trailing half of the MP
and it resembles an on-off switching of that component with little
evidence for a phase drift.  A similar modulation is also just visible
in the leading half of the MP and in the IP, although it is much
weaker.  Remarkably, the modulation of the IP appears to be more or
less in phase with that of the MP (see for instance pulses 200, 210
and 220).

\begin{figure}
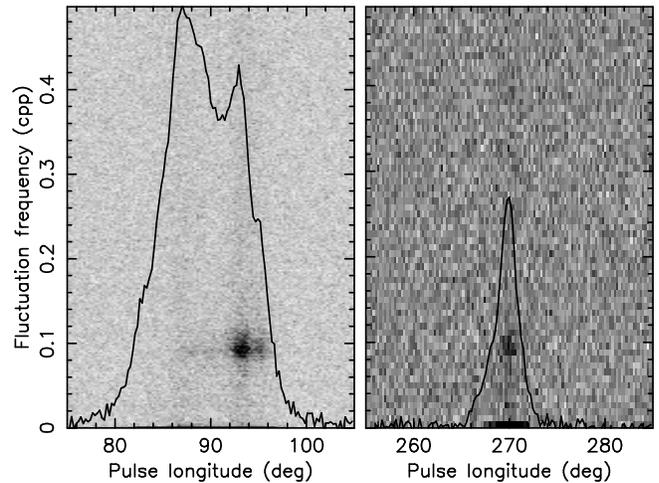

\begin{center}
\resizebox{!}{0.73\hsize}{\includegraphics[angle=0,trim=0 0 0 0,clip=true]{6957f4a.ps}}
\resizebox{!}{0.73\hsize}{\includegraphics[angle=0,trim=0 0 0 0,clip=true]{6957f4b.ps}}
\end{center}
\caption{\label{spectra}The LRFS (grayscale) of the 
2005L1 observation for the pulse longitude range around the MP (left)
and IP (right). The vertical scale of the LRFSs are in
cycles per period and the resolution is chosen differently for the MP
and IP to increase the contrast. Superimposed is the shape of the
pulse profile, which is normalized to have a peak intensity equal to
0.5. Notice that both the MP and IP show modulation with the same
periodicity.}
\end{figure}

The presence of an intensity modulation of the single pulses is
confirmed by calculating the Longitude Resolved Fluctuation Spectrum
(LRFS; \citealt{bac70b}, \citealt{es02}). The modulation is clearly
detected and indeed the periodicity found in the MP and IP are very
similar (Fig. \ref{spectra}).  For the observation shown we find
$P_3=10.4\pm0.3P$, which is consistent with the result of
\cite{wes06,wse07}.

Notice also that the modulation feature in the LRFS of the MP shows
horizontal structure.  At pulse longitude \degrees{95} there
is a distinct feature in the LRFS, which confirms the third emission
component which was also seen in the total intensity and linear
polarization profile (Fig. \ref{polprofile}).

\subsection{The phase locking of the modulation patterns}

\begin{figure}
\begin{center}
\resizebox{0.99\hsize}{!}{\includegraphics[angle=0,trim=0 0 0 0,clip=true]{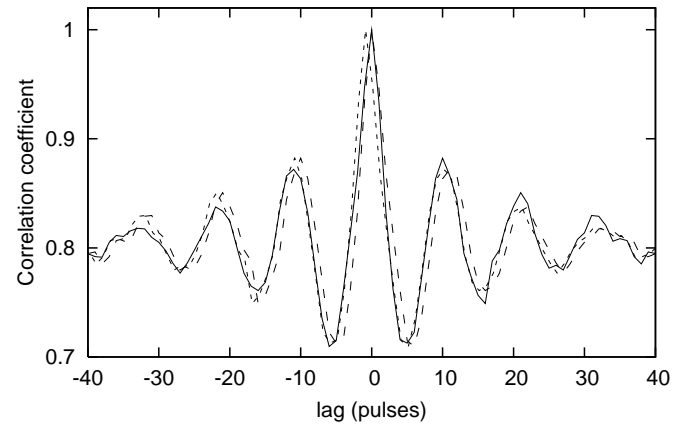}}
\end{center}
\caption{\label{correlation}The correlation between the pulse
intensities of the IP and that of the trailing peak of the MP for the
almost 19,000 pulses of the 2005L1 observation (solid
line). The correlation is normalized to the peak value and a
positive lag means that the MP signal lags that of the IP. The dashed
line shows the time reversed correlation and the dotted line shows the
time reversed correlation that is shifted with a lag of $-0.85P$.}
\end{figure}

\begin{figure*}
\begin{center}
\resizebox{!}{0.43\hsize}{\includegraphics[angle=0,trim=0 0 0 0,clip=true]{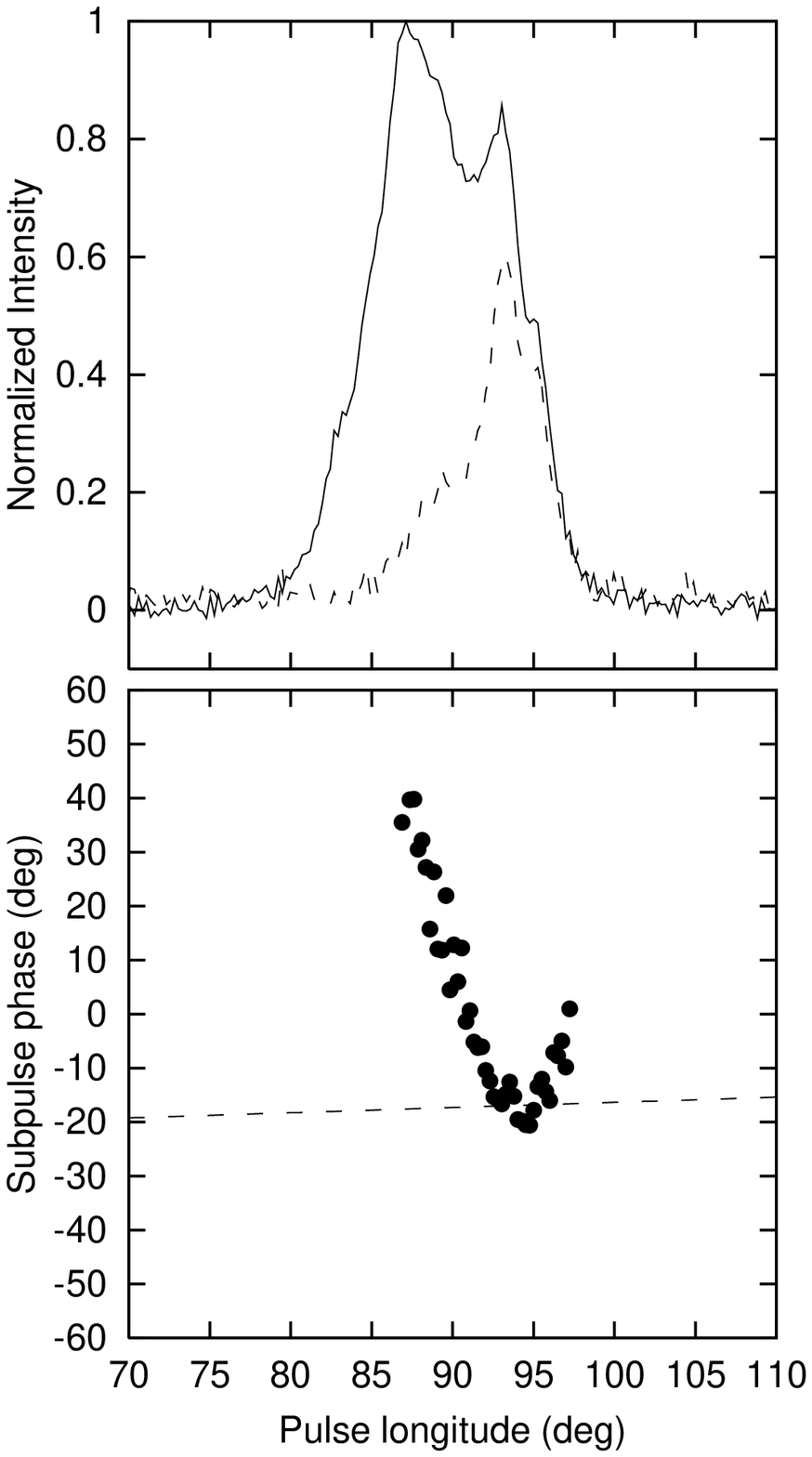}}
\resizebox{!}{0.43\hsize}{\includegraphics[angle=0,trim=15 0 0 0,clip=true]{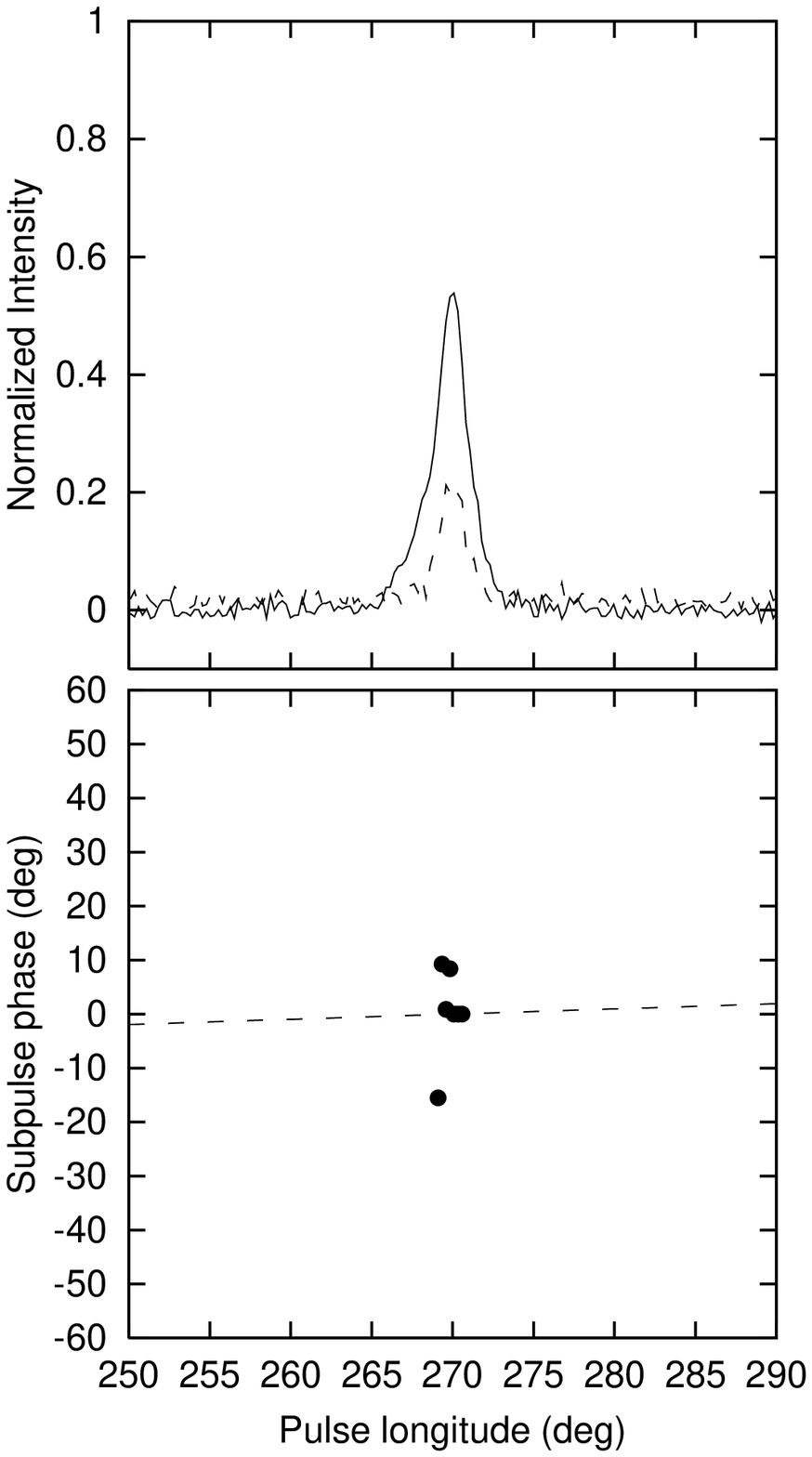}}
\hspace{7mm}
\resizebox{!}{0.43\hsize}{\includegraphics[angle=0,trim=0 0 0 0,clip=true]{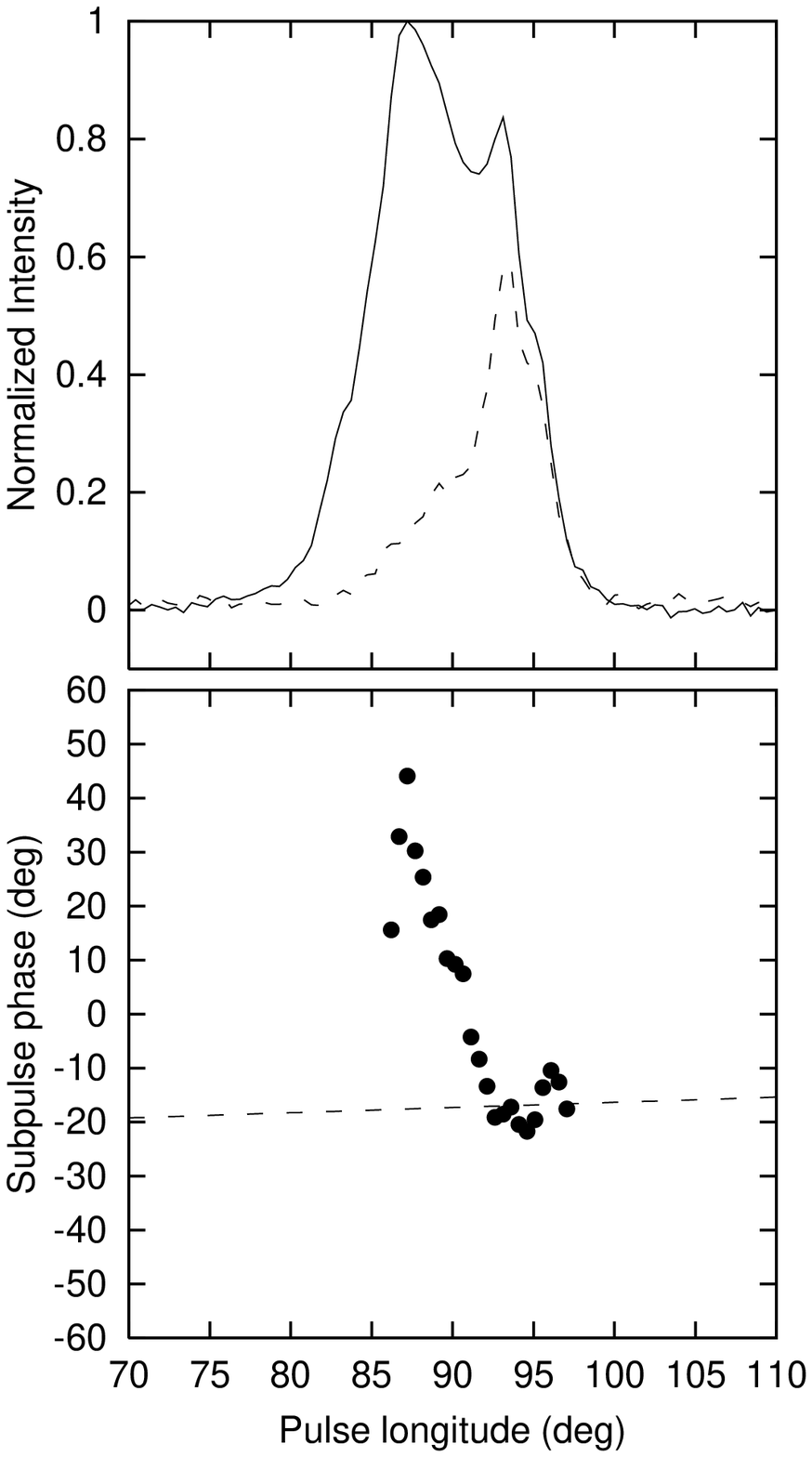}}
\resizebox{!}{0.43\hsize}{\includegraphics[angle=0,trim=15 0 0 0,clip=true]{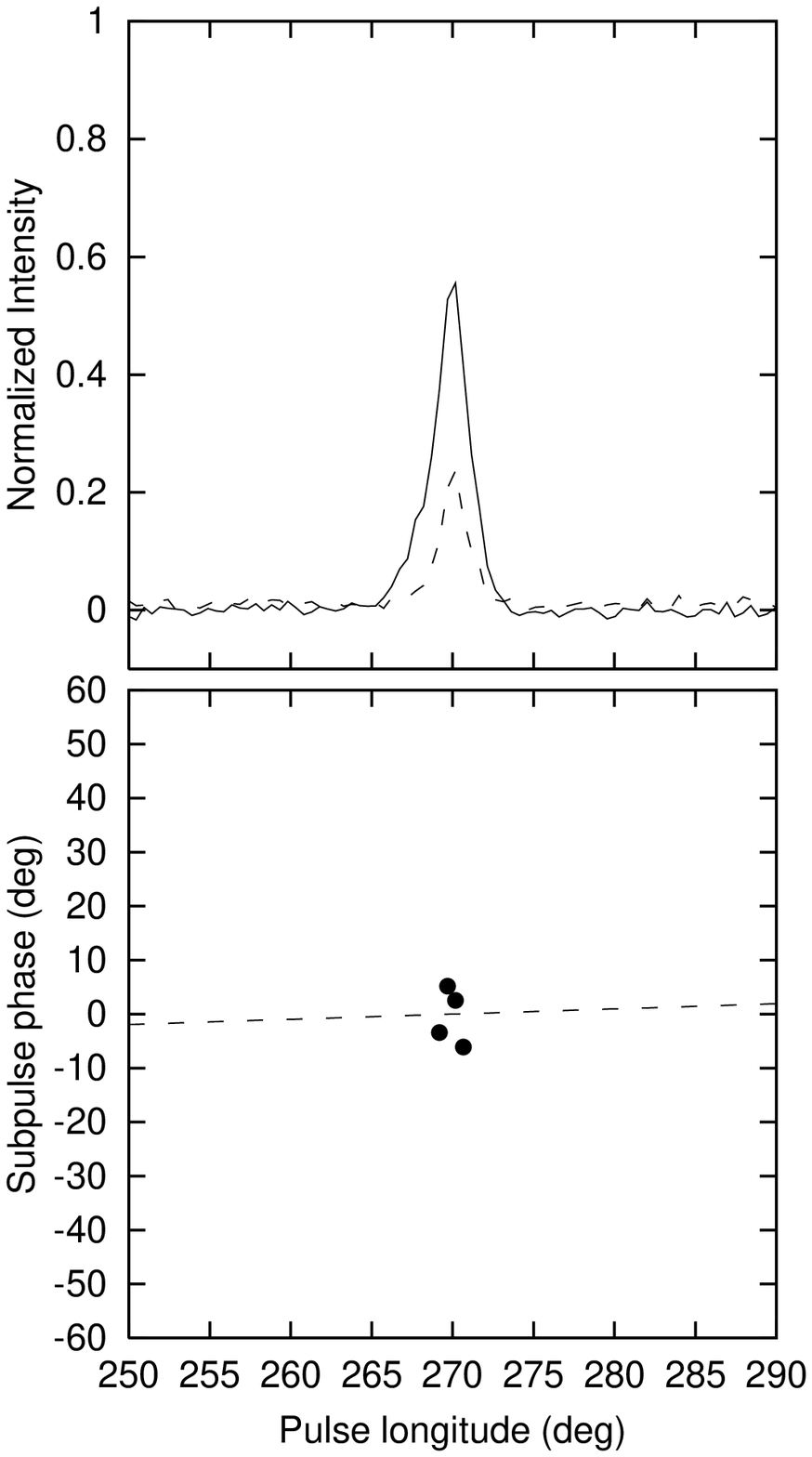}}
\end{center}
\caption{\label{tracks}The left two figures are the 2005L1 observation
and the right two figures the 2003L observation. The pulse longitude
ranges around the MP and IP are plotted separately. The top panels
show the normalized pulse profile (solid line) and the fraction of the
total intensity that is in the modulation (dashed line). The bottom
panels show the longitude-resolved subpulse phase for the
pulse longitudes where the modulation is detected above $3\sigma$. The
subpulse phase of the IP is set to zero. The subpulse phase at
different longitudes is intrinsically in phase for points on the
dashed line.}  
\end{figure*}

Because both the MP and IP show a reasonably coherent modulation with
possibly identical $P_3$ values there is an unique opportunity to
explore the subpulse phase connection between the two components. To
investigate this the single pulse intensities in the pulse longitude
range of the trailing half of the MP are integrated, giving the
strength of that component for each pulse. The same is done for the
IP. These arrays of integrated energies of the MP and IP are
correlated with each other (solid line in Fig. \ref{correlation}). The
$P_3\simeq10P$ modulation is clearly present in the cross correlation
function, which clearly confirms that both the MP and IP show a similar
modulation pattern.

The periodicity in the solid line of Fig. \ref{correlation} is not
only seen at a delay of $\pm10P$, but also at higher harmonics. This
is astonishing, because it demonstrates that the phase
difference between the modulation pattern of the MP and IP is not
significantly changed during the 19,000 pulses that were recorded. So
not only the $P_3$ value of the MP and IP are identical, but
apparently the modulation patterns are phase locked over long
timescales. This result could in principle be explained in two
different ways:
\begin{enumerate}
\item The modulation of both the MP and IP are perfectly 
stable with strictly identical $P_3$ values. 
\item The periodicity of the modulation patterns of the MP 
and IP vary slightly during the observation, but these variations are
identical (and occur simultaneously) in the MP and IP.
\end{enumerate}

The modulation feature in the LRFS (Fig. \ref{spectra}) 
shows vertical broadening, which means that the modulation frequency
 is varying slightly during the observation. This
interpretation could be further supported by the absence of any
correlation between the pulse energies of the MP during the first half
of the observation with the pulse energies of the IP during the second
half of the observation. This  clearly means that the first
option cannot be valid. There must be some mechanism operating in PSR
B1702--19 that keeps the phases of the modulation patterns of the MP
and IP locked.

A key result is that the correlation in Fig. \ref{correlation} is not
entirely symmetric under time reversal (compare the solid and dashed
line). If the time reversed correlation is shifted by a lag of
$-0.85P$, it matches the solid line with high precision. This
indicates that the modulation pattern of the MP lags that of the IP
pulse by $0.43P$. It can indeed be seen by eye in Fig.
\ref{pulsestack} that the modulation of the MP and IP are
roughly in phase. The measured phase difference of $0.43P$ is very
close to the $0.5P$ that  is expected because the IP emission
is observed half a stellar rotation later.
This means that the modulation patterns emitted by the MP and IP are,
correcting for this geometric delay, {\em intrinsically in phase}.

To investigate the intensity modulation in more detail, the
longitude-resolved subpulse phase and the strength of the modulation
is calculated for two observations (Fig. \ref{tracks}) using
the method developed by \cite{es02} and \cite{esv03}.
The top panels confirm that the trailing half of the MP and
the IP are highly modulated, while the leading component of the MP is
largely unmodulated, something that can also be seen directly in the
pulse-stack (Fig. \ref{pulsestack}). The 2006P observation 
(not plotted) also shows the modulation strongest in the
trailing half of the MP, showing the similarities of the modulation at
both frequencies.

The lower panels of Fig. \ref{tracks} show the subpulse phase tracks
of the modulation. For non-curved driftbands one expects a linear
relation between the subpulse phase and the pulse longitude. For
longitude-stationary subpulse modulation the subpulse phase is
independent of the pulse longitude (i.e. a horizontal relation).
The vertical phase scale on the left of the bottom panels is
such that \degrees{360} degrees corresponds to a full cycle $P_3$.
The absolute value of the subpulse phase has no direct
meaning, but differences in subpulse phase are interesting.

It can be seen in Fig. \ref{tracks} that the drifting
subpulse pattern for the MP has a complicated shape. In the trailing
half of the MP the subpulse phase is more or less independent of pulse
longitude, so the modulation is longitude stationary. However, between
the peaks of the MP the subpulse phase shows an inclined track,
indicating that the subpulses drift toward the leading peak
of the profile. This effect is quite rapid (it takes only about
\degrees{60} in subpulse phase, or about $1.7P$ before the emission
drifts off the pulse window) and can therefore only just be seen by
eye in the pulse-stack (Fig. \ref{pulsestack}). This result confirms
the presence of drifting subpulses reported by
\cite{wes06,wse07}.
Notice also that there is some indication that subpulses also drift
toward the extreme trailing edge of the MP, especially during the
2005L1 observation.

The subpulse phase of the IP is set to zero in Fig. \ref{tracks}.
Remarkably, not only is the shape of the subpulse phase
tracks very similar in the 2005 and 2003 observation, but also the
phase difference between the MP and IP is, within the measurement
uncertainties, identical in the two observations. This shows that {\em
the modulation patterns of the MP and IP are phase locked over a
period of at least several years}.

If the modulation pattern is intrinsically in phase at two
pulse longitudes then we expect the observed subpulse phases to be 
different, because the emission is received at two different
times. This geometrical delay is indicated by the dashed line in
Fig. \ref{tracks}, which has a slope of $P/P_3=1/10.4$ degrees per
degree.
In both observations the modulation of the IP is intrinsically in
phase with the second component of the MP, the two components of the
emission which are highly modulated.

\subsection{Intensity correlation}

\begin{figure}
\begin{center}
\resizebox{0.99\hsize}{!}{\includegraphics[angle=0,trim=0 0 0 0,clip=true]{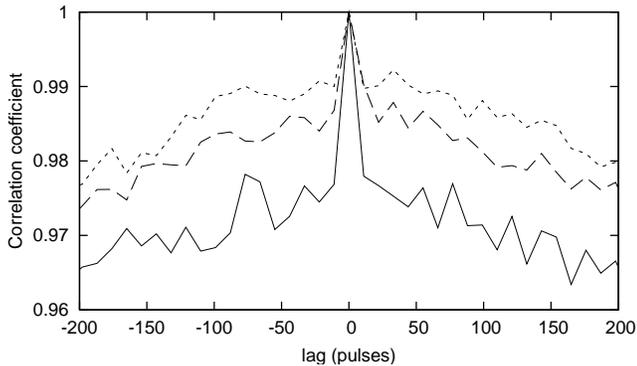}}
\end{center}
\caption{\label{correlationBlocks}The correlations between the integrated
pulse energies over 11 periods of the trailing peak of the MP with the
IP (solid line), leading peak of the MP with the IP (dashed line) and
between both halves of the MP (dotted line). In each case the
correlation coefficient is normalized to its peak value.}
\end{figure}

To further examine the degree of synchrony between the modulated
emission at the MP and IP, we explored the possibility that in
addition to a phase correlation between the various components of the
emission, there is also an intensity correlation. To investigate this,
each 11 successive pulses in the pulse-stack are summed  to
average out the $10.4P$ periodicity in the intensity modulation. This
is confirmed by calculating the LRFS of this sequence of integrated
pulse profiles. Apart from a $\sim5000P$ quasi-periodicity,
which is caused by the interstellar scintillation, no periodicities
remain.

The cross correlations of the integrated energies between different
parts of the profile are shown in Fig. \ref{correlationBlocks}. All
the correlations show a peak at a lag of zero superimposed on a much
slower decrease away from zero lag. The latter is caused by
interstellar scintillation. The solid line in
Fig. \ref{correlationBlocks} clearly shows that there is not only a
phase lock of the modulation patterns between the trailing part of the
MP with the IP, but also the intensity of the components are
correlated. Notice also that the cross correlation between the two
halves of the MP is weaker than the correlation between the leading
half of the profile and the IP. This could be related to the fact that
the proportion of the unmodulated emission in the IP is higher than
that in the trailing component of the MP (see Fig. \ref{tracks}). If
not only the intensity of the modulated part of the emission is
correlated, but also the intensity variations in the unmodulated part
of the emission, a stronger correlation is to be expected between the
leading half of the profile and the IP than between the two halves of
the MP.

\section{\label{SctPA}The viewing geometry}
\begin{figure*}[tb]
\begin{center}
\resizebox{0.45\hsize}{!}{\includegraphics[angle=0,trim=0 0 0 0,clip=true]{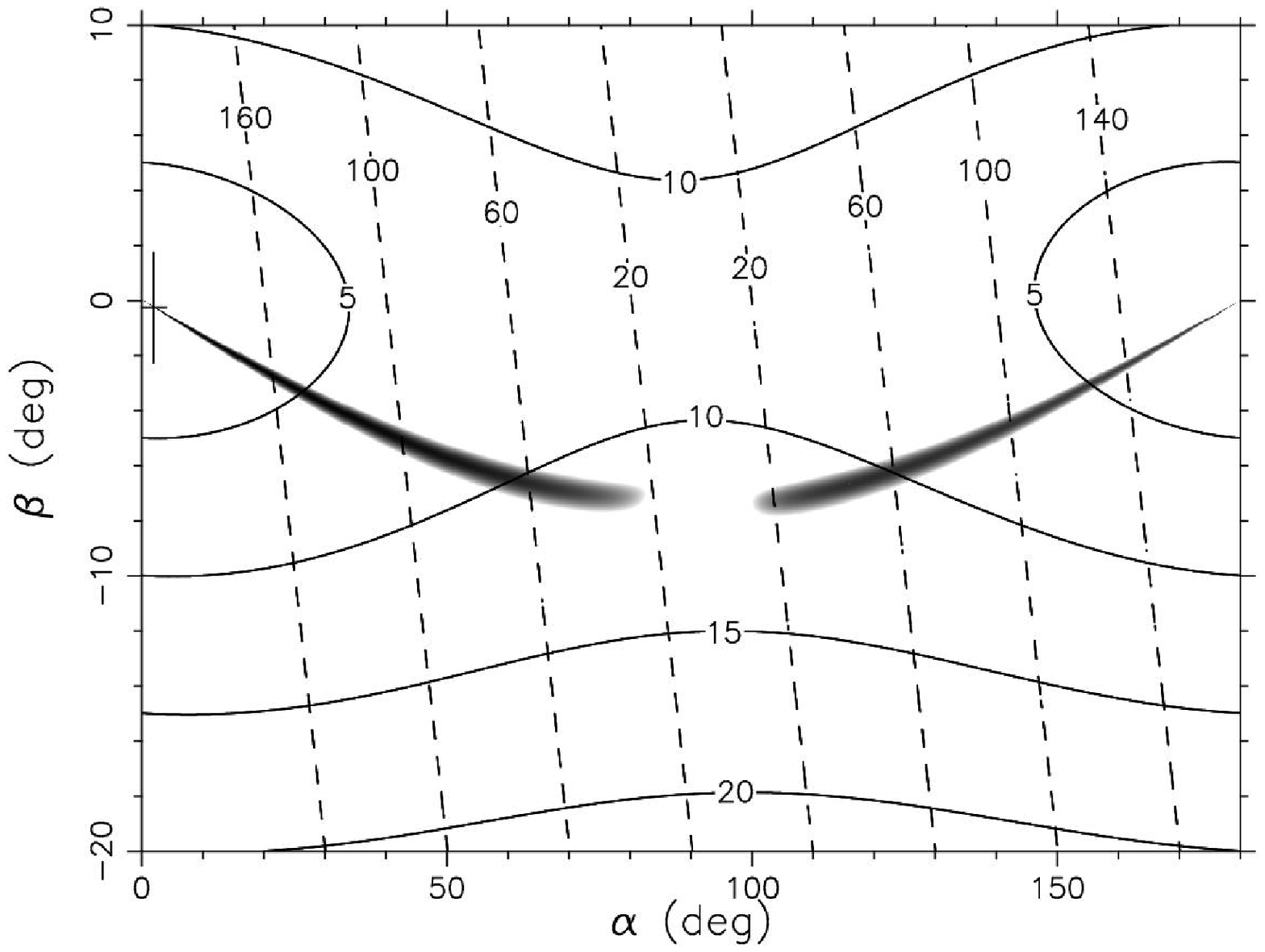}}
\hspace*{1cm}
\resizebox{0.45\hsize}{!}{\includegraphics[angle=0,trim=0 0 0 0,clip=true]{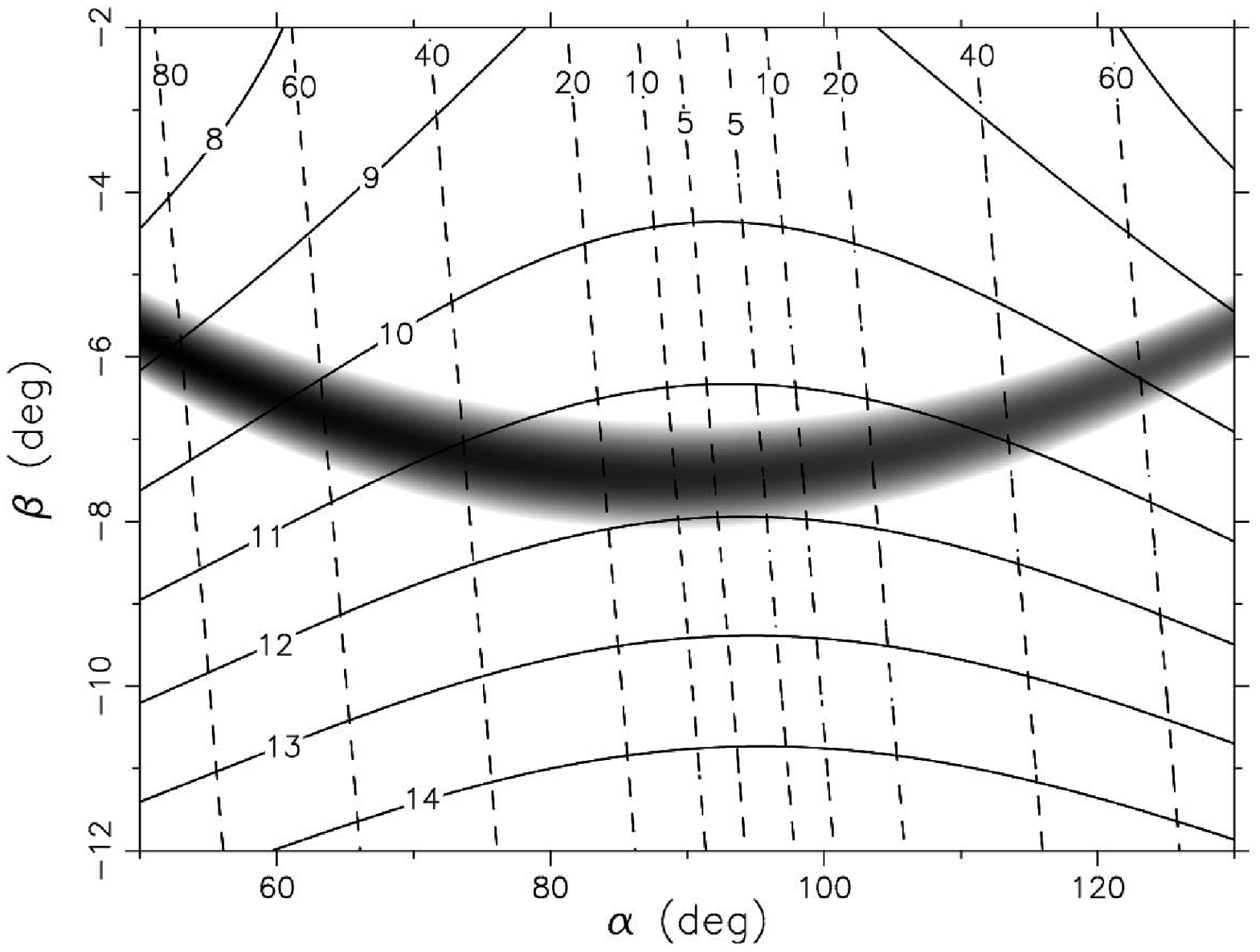}}
\caption{\label{twopole}{\bf Left:} The $\chi^2$ distribution found by
fitting of the  entire PA-swing in the two-pole
scenario. White corresponds with $\chi^2=15$ and black with
$\chi^2=6$. The position of the cross (at
$\alpha=\degrees{1.07}$) indicates the best fit. The  solid
and  dashed contours show the derived beam radii from the
pulse width of the MP and IP. {\bf Right:} The right panel shows the
$\chi^2$ distribution when only the PA-swing of the MP is fitted.  The
plot is zoomed in around $\alpha=\degrees{90}$.}
\end{center}
\end{figure*}

\subsection{The PA-swing}

When the emission of the pulsar is linearly polarized the Position
Angle (PA) can be measured and the PA-swing is plotted in the bottom
panel of Fig. \ref{polprofile}. The shape of PA-swings are typically
S-shaped, which is interpreted in the rotating vector model (RVM;
\citealt{rc69a}) as the projection of dipolar fieldlines on the plane
of the sky. Its precise shape depends on the orientation of the pulsar
with respect to the line of sight. In particular it depends on the
angle $\alpha$ between the magnetic axis and the rotation axis and the
impact parameter $\beta$, which is the angle between the line of sight
and magnetic axis at the position of the closest approach. For a line
of sight between magnetic and rotation axis, $\beta$ is negative if
$\alpha < \degrees{90}$ and positive if $\alpha > \degrees{90}$.
Another angle that is often referred to in the literature is
$\xi=\alpha+\beta$, which is the angle between the line of sight and
the rotation axis.

The PA-swing of the MP is quite steep and smooth and can be understood
in terms of the RVM. However, the PA-swing in the IP has a U-shape,
something that is not expected in the RVM. Also \cite{gl98} note that
the PA of the IP has an essentially flat rotation at the higher
frequencies, while it is more U-shaped at lower
frequencies. It is clear that the shape of the PA-swing of the IP
cannot be used to constrain the geometry. However, if the deviations
from the RVM are relatively small then the average magnitude of the PA
in the IP still obeys the RVM and contains useful information.

The aim in this section is to derive the geometry ($\alpha$ and
$\beta$) from the shape of the PA-swing.
The basic approach is to do a grid search over all possible
combinations of $\alpha$ and $\beta$ and for each combination the
amoeba search algorithm (\citealt{ptv+92}) is used to minimize the
$\chi^2$ (which is a measure for the discrepancy between the measured
and modeled PA-swing). The measurement uncertainties of the
PA are taken into account in the $\chi^2$. The fit parameters are the
pulse longitude of the magnetic axis and the PA at that position.

The fact that PSR B1702--19 has an IP is an important
additional constraint to the allowed geometries. IPs can be
interpreted in different ways and three different scenarios are
explored in this section. One of them corresponds to a perpendicular
rotator ($\alpha\simeq\degrees{90}$) and two to an aligned rotator
($\alpha\simeq\degrees{0}$).  The most straightforward interpretation
(\citealt{rl68}) 
is a perpendicular rotator where one pole corresponds to the MP and
the other to the IP. The other two scenarios involve emission from a
single pole. We examine each of them in the light of our polarisation
measurements.

\subsection{\label{twopolemodelfit}Two-pole model}

In the two-pole model the line of sight probes the regions near both
magnetic poles ($\alpha\simeq\degrees{90}$). It is important to note
that for the shape of the PA-swing it does not matter if we only see
one or two poles. This means that the standard RVM can be used for
both one and two-pole models.

In the left panel of Fig. \ref{twopole} the resulting $\chi^2$
distribution is shown (in grayscale) for the two-pole model. The best
fit has a reduced $\chi^2\simeq 6$. The $\chi^2$ distribution is
basically a single arc that covers the whole $\alpha$ range, so
$\alpha$ is not well constrained by the shape of the
PA-swing. Solutions near $\alpha=\degrees{90}$ appear to be less
favoured, which can be understood because the PA-swing of the IP is
 very flat and not described well by the RVM. When only the
PA-swing of the MP is fitted,  a $\chi^2$ distribution is
obtained that allows solutions near $\alpha=\degrees{90}$ (right panel
of Fig. \ref{twopole}). The PA-swing in the IP is much flatter than
that of the MP and it is not even clear if the PA-swing is rising or
declining in the IP. This means that the PA-swing of the IP does not
constrain the fits much, except that solutions near
$\alpha=\degrees{90}$ are less favoured.

The fact that the PA-swing of the IP appears to be U-shaped
clearly shows that it deviates significantly from anything that would
be consistent in any detail with the RVM. Nevertheless, the RVM
predicts that the PA should have similar values at the longitudes of
the MP and IP, consistent with the observations.  Therefore it is not
obvious that the flat PA-swing of the IP implies that $\alpha$ cannot
be close to \degrees{90}. A roughly orthogonal orientation of the
magnetic axis would be consistent with the reports by
\cite{vdhm97,lm88,kuz89}.  \cite{gl98} find evidence that there is a
\degrees{90} discontinuity in the PA-swing at the leading edge of the
MP, suggesting the presence of two Orthogonal Polarization Modes
(OPMs). However, it seems unlikely that the MP and IP are dominated by
different OPMs, because they show a similar PA.

The geometry of the system can be further constrained by considering
the beam radii $\rho$ of the MP and IP. The beam radius $\rho$ can be
derived from the measured pulse width $W$ (e.g. \citealt{lk05}) and it
depends on the geometry as follows:
\begin{equation}
\label{cos_rho}
\cos\rho=\cos\alpha\cos\xi+\sin\alpha\sin\xi\cos\left(\frac{W}{2}\right).
\end{equation}
For the IP the $W$ is the width of the IP and the sign of $\cos\alpha$
changes\footnote{The geometry of the IP (indicated with
primes) is related to that of the MP (without primes) as follows:
$\alpha^\prime = \degrees{180} - \alpha$, $\beta^\prime =
\beta+2\alpha-\degrees{180}$ and $\xi^\prime=\xi$.}.
The measured widths $W$ for the MP and IP are \degrees{18} and
\degrees{7} respectively (the full widths as well as can be accurately
determined given the $S/N$). The corresponding beam radii are overlaid
as contours in the plots of Fig. \ref{twopole}.  
As expected in a two-pole scenario $\alpha$ values close to
$\degrees{90}$ are favoured because only then the beam radii of the MP
and IP are similar to each other and have acceptable values.

\begin{figure}
\begin{center}
\resizebox{0.99\hsize}{!}{\includegraphics[angle=0,trim=0 0 0 0,clip=true]{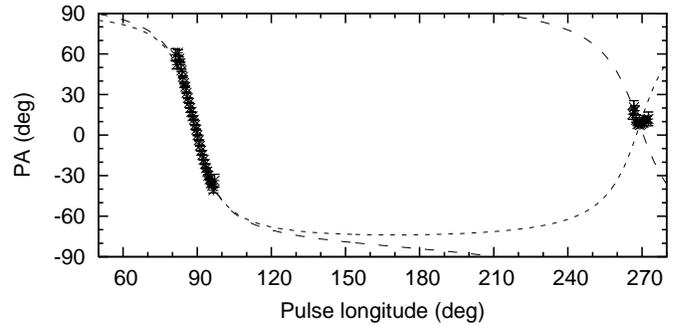}}
\caption{\label{fit88}The best fit of the PA-swing for the  
$\alpha=\degrees{88}$ and $\beta=\degrees{-7.5}$ solution (dashed
line) and the $\alpha=\degrees{99}$ and $\beta=\degrees{-7.5}$
solution (dotted line). These solutions both have equal beam
widths for the MP and IP ($\rho\simeq\degrees{12}$).}
\end{center}
\end{figure}

It seems that the IP is roughly described by the RVM in the two-pole
model, but is distorted by an unknown process. Therefore it makes
sense to only use the MP to determine the geometry using the RVM
(right panel of Fig. \ref{twopole}). It can be seen that for
$\alpha\simeq\degrees{88}$ and $\alpha\simeq\degrees{99}$ the beam
radii of the MP and IP are equal to each other
($\rho\simeq\degrees{12}$). These two solutions are plotted in
Fig. \ref{fit88}. Both solutions describe the PA-swing of the MP
equally well, but they are very different for the IP.

Assuming that the radius of the beam is set by the last open
fieldline, $\rho$ can be linked to the emission height
(e.g. \citealt{lk05}) and can be approximated for $\rho
\lesssim\degrees{30}$ to
\begin{equation}
\label{emissionheight}
\rho=\degrees{1.24}\left(\frac{r_\mathrm{em}}{10\;\mathrm{km}}\right)^{1/2}\left(\frac{P}{1\mathrm{s}}\right)^{-1/2}.
\end{equation}
Thus for a beam radius of \degrees{12} and $P=0.299$ seconds the
emission height is $r_\mathrm{em}=280$ km.

\subsection{Single wide cone model}

In the {\em single wide-cone} model (\citealt{ml77}), the pulse
profile is explained as a double peaked profile with a very large
component separation. Such a model works best, in general, for a more
aligned rotator than expected for a two-pole model.  The MP and IP of
PSR B1259--63 are for instance interpreted as a wide cone
(\citealt{mj95}).
Wide cones are especially expected for (young) pulsars with
short-periods, because their beams are wider. In this interpretation a
component separation very close to \degrees{180} has to be considered
to be a coincidence and the absence of a ``bridge'' of emission
between the components also makes this interpretation less likely.

In the wide cone model the position of the magnetic axis ($\phi_0$)
should be located between the MP and IP. Using a grid search over all
possible combinations of $\alpha$ and $\beta$ with the constraint that
$\phi_0$ should be between pulse longitude
\degrees{125} and \degrees{215} revealed that the lowest reduced
$\chi^2$ found is almost 1000. This means that the best possible fit
clearly fails to describe the observed PA-swing. The reason is that
the PA-swing in the MP is too steep compared with the model. A steep
PA-swing implies that the magnetic axis should be close to the
corresponding pulse longitude, which is not allowed in the wide cone
model. Indeed, the least worst fit is found to have the magnetic pole
placed right at the border of the allowed range.

Similar to the procedure for the two-cone model it is interesting to
consider the possibility that the IP is in a different OPM.
However, the fits turn out even worse when a jump is included in the
model, because even flatter PA-swings are predicted in such a case.
So in both the single wide-cone model and the two-pole model it seems
unlikely that the MP and IP have a different OPM.

It is clear that a single wide-cone model cannot explain the observed
PA-swing, because
the steepness of the PA-swing of the MP can only be explained by the
RVM when the magnetic axis is close to the MP. Indeed the PA-swing of
PSR B1259--63, for which a wide cone is proposed (\citealt{mj95}), is much flatter.

\subsection{Inner and outer beam model}

\begin{figure}
\begin{center}
\resizebox{0.6\hsize}{!}{\includegraphics[angle=0,trim=0 0 0 0,clip=true]{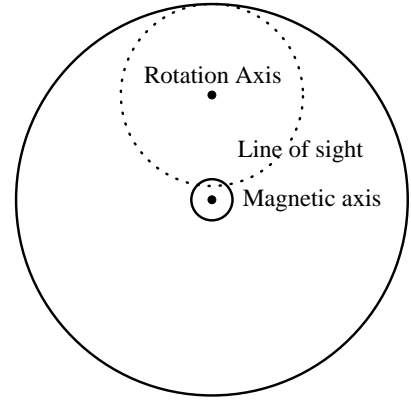}}
\caption{\label{twoconesgeometry}The inner and outer beam 
interpretation for the best fit of the PA-swing with a small $\alpha$.
The figure is drawn to scale. The MP is produced by the inner beam and
the IP by the outer beam.}
\end{center}
\end{figure}

An alternative single pole model, also for a nearly aligned
rotator, is proposed by \cite{gil85}. 
This model
has to be considered as a valid alternative since it also creates a frequency independent
\degrees{180} separation naturally. It consists of concentric inner
and outer cones surrounding a single magnetic pole which is close to
alignment with the rotation axis (Fig. \ref{twoconesgeometry}). In
this model either the MP or the IP component must lie where the line
of sight passes close to the magnetic pole, so we can apply the
results of the two-pole model (Fig. \ref{twopole}). The solutions with
lowest $\chi^2$ are found for very small $\alpha$ because in this case
the PA-swing in the IP is expected to be slightly flatter. The best
fit has $\alpha=\degrees{1.07}$ and $\beta=-\degrees{0.14}$ and the MP
is associated with the inner cone.

This interpretation could in principle also explain why the MP-IP
separation is close to \degrees{180} (and independent of
frequency). However, it seems that the radius of especially the outer
cone must be chosen exactly right to avoid the width of the
IP becoming huge. Also, the width of the IP could be
expected to be highly dependent on frequency, something that is not
observed.

The beam radius of the inner cone obeys Eq. \ref{cos_rho}, while the
beam radius of the outer cone can be calculated with the
same equation with $\beta$ replaced by $-2\alpha-\beta$ (and $W$ is
the width of the IP)\footnote{This follows directly from
Fig. \ref{twoconesgeometry}.}. 
This means that to explain the measured widths of the MP and IP
(respectively \degrees{18} and \degrees{7}) the ratio between the beam
radius of the outer and inner cone for the best fit $\alpha$ and
$\beta$ values is 9.5.
Note that $\beta$ is highly constrained for a given
$\alpha$ such that there is a more or less linear relation between
$\alpha$ and $\beta$. This means that for all choices of $\alpha$ (as
long as it is small) the ratio between the inner and outer cone should
also be 9.5.

From Eq. \ref{emissionheight} it is clear that the radius of the beam
cannot be smaller than \degrees{1.24}, which rules out the best fit
with $\alpha=\degrees{1.07}$. 
In the case that the inner and outer beams are produced on the same
field lines, but at different heights, the emission height of the
outer component should be $(9.5)^2$ times larger than that of the
inner beam. In the extreme case that the MP is produced at the surface
of the star, the emission height of the outer component should be
larger than 900 km.

Retardation and aberration will cause emission from higher
in the magnetosphere to arrive earlier at the observer and a difference
$\Delta h_\mathrm{em}$ in emission height corresponds to a pulse
longitude shift of 
\begin{equation}
\label{Eqtwopoleshift}
\Delta \phi = \frac{4\pi\Delta h_\mathrm{em}}{cP},
\end{equation}
where $c$ is the speed of light.  A differential emission height of at
least 900 km corresponds to a combined aberration and retardation
effect of at least \degrees{7} in pulse longitude, which is
independent of geometry (\citealt{drh04}). It therefore appears that
even in the most extreme case the MP-IP separation could be expected
to deviate strongly from
\degrees{180}, making this model implausible. The light cylinder
radius is $1.4\times10^{4}$ km, implying that in this model the
emission height of the inner cone must be less than $155$ km.

\section{\label{SctDisc1702}Discussion}

In the preceding sections we have established the basic radio emission
properties of PSR B1702--19. It is now appropriate to organise the
data in an attempt to assemble a coherent model.

Two major conclusions have been reached. First, that the
polarisation data is best described by a perpendicular rotator --
i.e. by two oppositely directed rotating vectors swept alternately
through the observer's sightline. These vectors could come from
opposite magnetic poles of a dipole, but we cannot rule out {\em
a~priori} that these two vectors are present on the same side of the
pulsar, e.g. produced by upflow and downflow on the selfsame polar
fieldlines at a single pole (the {\em bidirectional model}). However,
this model is asymmetric and at typical emission heights the combined
effects of aberration and retardation would displace their
corresponding profile images from an exact
\degrees{180} separation.

Second, we have conclusively shown that the modulation in the MP and
IP are locked into a single system, whether the pulsar's emission is
examined over the long or short term. The measurable weak fluctuations
in the modulation periodicity occur simultaneously at both longitudes,
and this locking is maintained over tens of thousands of pulses, and
even years (Figs. \ref{correlation} and \ref{tracks}). Further, we
find that the intensities of the modulated emission in the MP and IP
(and not only its precise repetition pattern) are correlated
(Fig. \ref{correlationBlocks}). A final compelling argument, on the
shortest timescale available, is that a cross-correlation between the
trailing component of the the MP and the following IP consistently
yield a phase-delay in the modulations close to 0.5P, exactly what is
to be expected if both are $\emph{intrinsically}$ in phase
(Figs. \ref{correlation} and \ref{tracks}).

We will consider the implications of the two-pole and bidirectional
models using only the minimum of physical assumptions. The effects of
aberration and retardation are incorporated, but more complicating
effects such as gravitational bending and refraction of the emission
beam are ignored. This is not because we believe that these effects
are unimportant, but because it is very difficult to accurately
estimate them, and because it is important to establish whether a
model can be constructed without their aid.

\begin{figure}
\begin{center}
\resizebox{0.99\hsize}{!}{\includegraphics[angle=0,trim=0 0 0 0,clip=true]{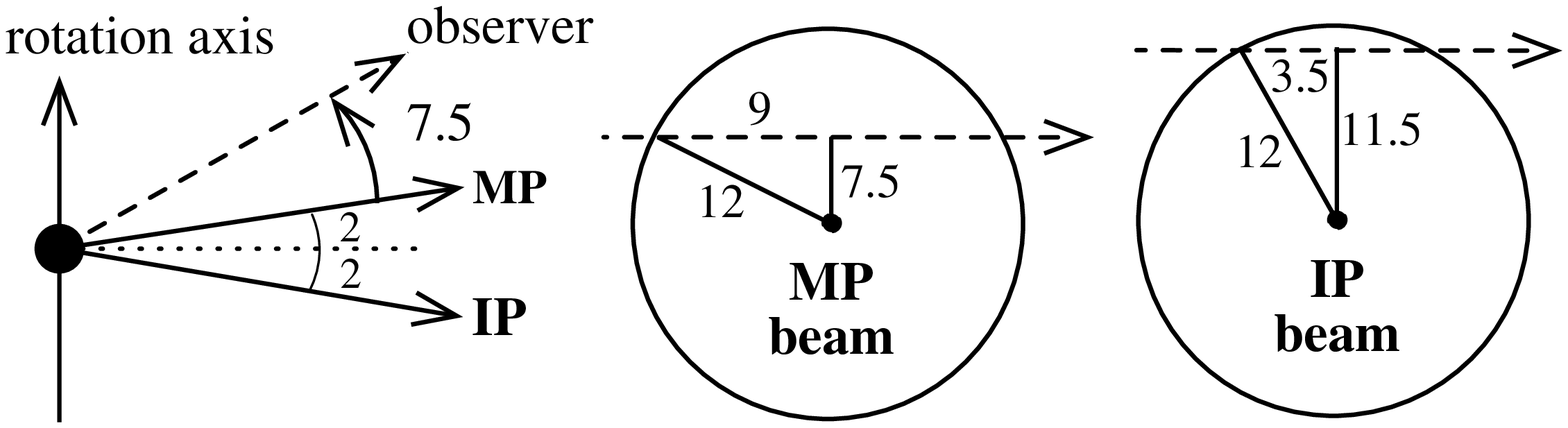}}\\
\vspace*{5mm}
\resizebox{0.99\hsize}{!}{\includegraphics[angle=0,trim=0 0 0 0,clip=true]{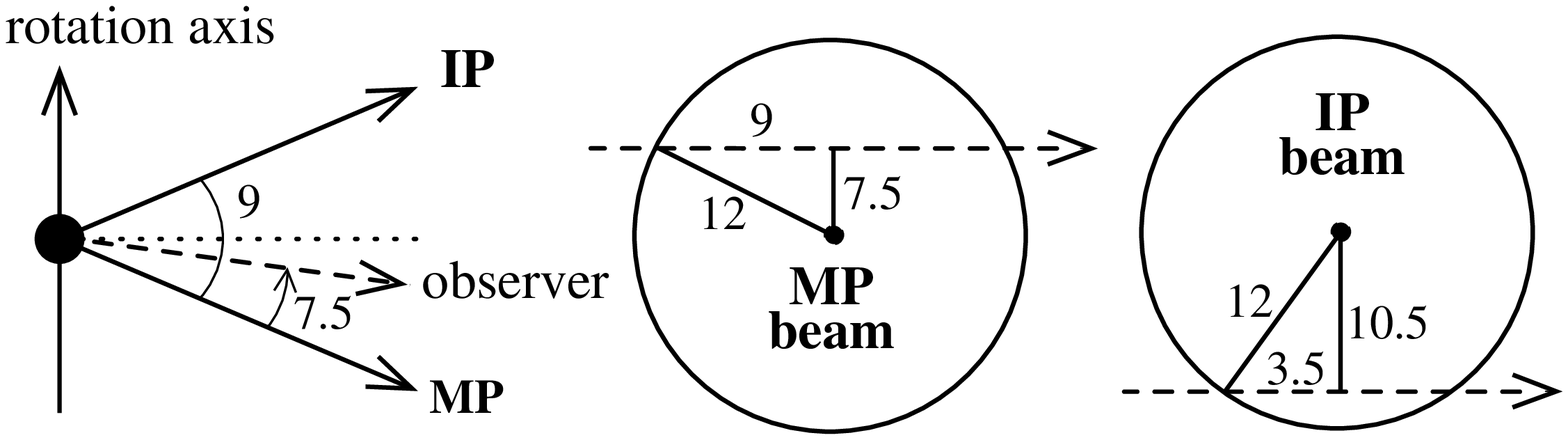}}\\
\end{center}
\caption{\label{fig1}The observer's MP/IP views of the pulsar's beam for the 
case $\alpha=\degrees{88}$ (top  figures) and
$\alpha=\degrees{99}$ (bottom figures). In the top
figures the sightline intersects the MP and IP on the same
side of the magnetic pole, but significantly separated. In the bottom
figures the sightline intersections of the MP and IP beams
are widely separated on opposite sides of the magnetic pole. 
The beam widths ($\rho\simeq\degrees{12}$), the measured half widths
of the MP and IP (respectively \degrees{9} and
\degrees{3.5}) and the impact angles $\beta$ are indicated as well.}
\end{figure}

\subsection{\label{SctDiscTwoPole}Two-pole model}

In a model where the modulated emission is coming from two opposing
poles, the simplest assumption is that PSR B1702--19 is a
near-perpendicular rotating dipole and that the two poles possess an
identical emission system (there is no evidence of one pole
``driving'' the other). One might object that the MP and IP profiles
manifestly differ in shape and proportion. However, neither the
observer's sightline nor the pulsar's beam is precisely perpendicular
to the rotation axis and, even if the beams at both poles are
identical, the observer's sightline will make differing
cuts. Fig. \ref{fig1} shows how this may occur based on the two fits
for the PA-swing from Fig. \ref{fit88}, which have equal beam
widths. Clearly, in both cases the observer has a ``fatter'' view of
the beam at the MP than at the IP, where the sightline merely grazes
it.

Differing sightlines may therefore explain the different profiles of
the MP and IP, but their observed synchrony has major implications for
the emission patterns of the beams. Unless, at any given moment, the
intensity of both beams is uniformly distributed, or possibly
ring-like, then there is no reason to expect synchrony between two
completely different sections of the beam. This is particularly true
for the fit with $\alpha=\degrees{99}$ (bottom panel of
Fig. \ref{fig1}), where the sightline intersects the beam on opposite
sides of the magnetic pole. If one believes that the beam intensity
has angular structure, whether arising from a circulating ``carousel''
(e.g. \citealt{rs75}) or varying random patches 
(\citealt{lm88}), it must be considered coincidental that we are able
to observe two regions of the beams which are so precisely in phase
with each other. This argument also applies when the MP modulation is
considered alone. Fig. \ref{spectra} suggests a conal structure in the
trailing second and third components  of the MP, and that
the modulation is occurring on both sides of the intersected cone. Yet
Fig. \ref{tracks} shows these regions to be in phase, which need not
be expected from an angularly structured emission beam.

Even if we allow this phase coincidence, then the inter-pole synchrony
is still physically difficult to understand. In the \cite{rs75} model
and its successors  (e.g. \citealt{gs00} and
\citealt{gmg03}) the ``sparking'' distribution is fixed by the
multipole structure on the surface of the neutron star surrounding the
pole, 
and there is no obvious reason why these should be the same at both
poles, and hence produce the same emission patterns. Even in the
empirical model of
\cite{wri03}, which requires feedback over large distances within the
magnetosphere for its operation and which therefore might be expected
to support inter-pole contact, it is hard to see why the system would
need to arrange itself so that the poles would be in exact phase
synchrony. Information travelling at the speed of light along a closed
magnetic fieldline from one pole would take at the very
least $0.44P$ to reach the other (\citealt{big90a}). This time is far
longer than the observed degree of phase locking, hence simultaneity
at both poles could not be enforced in this environment.  The
only way events at the two poles could be kept locally simultaneous
would be if information could travel through the body of the neutron
star or its surface. Possible physical mechanisms for this might be
oscillations of the star (e.g. \citealt{van80}, \citealt{cr04}), or
variations in its net electrical charge (e.g. \citealt{ml99}),
but no detailed model based on these ideas has been worked out.

However, it should be noted that the modulation of PSR
B1702--19 is not typical of that found in pulsars with coherent
drifting subpulses. The modulation is dominated by intensity changes,
rather than phase changes. These latter, though clearly discernible in
Fig. \ref{tracks}, amount to only \degrees{60} of phase across the
pulse window, and appear to have opposing drift directions on either
side of the centroid of the trailing component of the MP.
The rapid subpulse drift across the profile is also evident in
Fig. \ref{pulsestack}.
In the vast majority of pulsars with regular drift patterns 
(e.g. PSR B0809+74 which has a very similar $P_3$) the subpulses drift slowly through \degrees{360}
in subpulse phase in the pulse window and are mono-directional, even
when the drift-mode undergoes a change. This contrasting behaviour may
be due to the differing inclinations of the pulsar's magnetic axis
with respect to the rotation axis. Slow-drifting pulsars such as PSR
B0809+74 are often found to be relatively close to alignment
($\alpha<\degrees{10}$) (\citealt{lm88},
\citealt{ran93b}, \citealt{wri03}, \citealt{rw03}), while PSR B1702--19 is almost
perpendicular, a factor which is likely to lead to different electric
potential distributions and beam shapes close to the poles
(\citealt{rud76b}).

The overall structure of the MP is difficult to account for. The
trailing, highly modulated component, with its hint of conal structure
mentioned above, is in fact only \degrees{176} ahead of the IP
centroid, and the role of the largely unmodulated leading component is
puzzling. In the two-pole picture, the most natural explanation for
the leading component would be to consider it as an integral part of
the main beam, whose midpoint is 
separated from the IP by exactly \degrees{180} at both our observing
frequencies (see Fig. \ref{profiles_two_freq}). Nothing in the smooth
PA-swing across the entire MP suggests a different origin for this
component, and the absence of modulation at leading longitudes, while
mysterious, is also found in several pulsars with drifting emission
(\citealt{wes06,wse07}). It is however strange that its intensity is
so weakly coupled with that of the modulated component over the $P_3$
cycle (Fig. \ref{correlationBlocks}) and appears to be virtually
decoupled from the pulsar's modulation cycle. It might be possible to
claim that this component represents 'core' emission, which is
supported by the presence of strong circular polarisation which often
accompanies such a feature (\citealt{ran90}).
However, core components in highly-inclined pulsars are normally
narrow and are thought by some (\citealt{ran90}) to be formed close to
the neutron star surface. In any case, core components should, by
definition, form close to the magnetic axis, and there is nothing in
the RVM fits to suggest that our sightline passes close to the axis.
Perhaps the best that can be said is that the component does not have
any apparent conal characteristics.

\subsection{Bidirectional emission}

It has recently been suggested (\citealt{dzg05}) that at
least some of the MP emission in pulsars with IPs of nearly
\degrees{180} separation from the MP may be directed downwards towards
the neutron star and viewed from the opposite side as an IP. This
radical idea may be usefully applied to PSR B1702--19, since it offers a
natural way to explain the synchrony of the trailing component of the
MP and the IP by supposing that the source of the modulated emission
is viewed twice per stellar rotation. Whatever the physical nature of
the source, it would seem at a stroke to explain why the modulation is
exactly in phase over short and long timescales. However, as with the
two-pole model of the previous section, it does {\em not} explain why
the structure of the emission beam (Fig. \ref{fig1}) is so tightly in
phase, and we must assume that, whether viewed from front or back, the
emission must have a ring or uniform structure within the beam.

The resulting geometry of this bidirectional model differs
in significant points of detail from that of \cite{dzg05}, since these
authors associate different components of the pulse profile with the
reversible source in the pulsar they analyse (PSR B1822--09, see next
section). But it shares with that model a fundamental difference to
the classical two-pole model of Section \ref{SctDiscTwoPole}, in that
viewing a rotating vector at a fixed difference above the stellar
surface will inevitably entail aberration and retardation effects,
causing the viewer to see the back view of the vector slightly later
than \degrees{180} from its front view.

If we assume the midpoint of the MP occurs at the minimum
between its two components, then the bidirectional model cannot be
correct. This is because this point is precisely \degrees{180} from
the centroid of the IP, so it would imply no aberrational of
retardation effects. However, if we take an alternative view of the MP
based on the pulsar's subpulse behaviour rather than its integrated
profile we come to a different picture. Disregarding for the moment
the unmodulated leading component of the MP and focus on the
phase-locked modulated components, then we notice first of all that
the modulated second component of the MP is clearly an overlapping
double component (Fig. \ref{spectra}), which are seen to be in
phase-lock with the IP (Fig. \ref{tracks}). Therefore the natural
choice for the centroid of the modulated second component of the MP
would be the midpoint between the two modulated regions, which would
be the centre of the postulated conal ring of modulated emission.

\begin{figure}
\begin{center}
\resizebox{0.99\hsize}{!}{\includegraphics[angle=0,trim=0 0 0 5,clip=true]{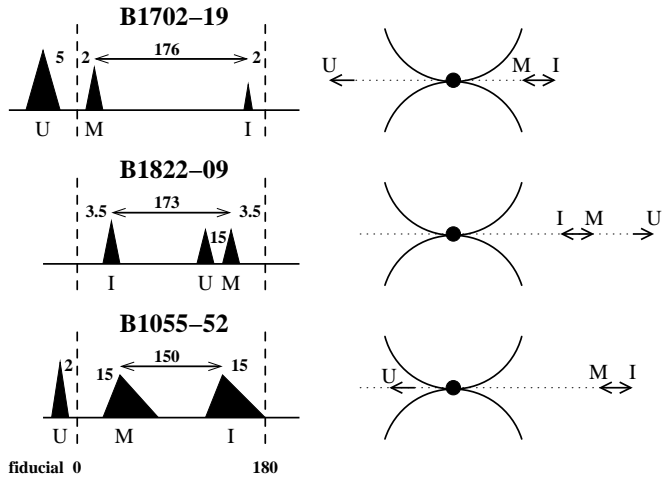}}
\end{center}
\caption{\label{fig2}The relative phase separations (in degrees) of 
PSRs B1702--19, B1822--09 and B1055--52, and their translations into a
bidirectional model. The fiducial points of each refers to the derived
longitude of the magnetic poles on the star at the instant they are
directed to the observer. Both the phase plots (left) and dipolar
models (right) are indicative and not to scale. The IP, the modulated
and unmodulated components are indicated by respectively I, M and U
(see text for estimated heights).
}
\end{figure}

This point is \degrees{176} ahead of the IP, a shift of
\degrees{4} from the MP's emission minimum. If this point and the IP
are to be considered as sightline sections across reverse images of
one another, then the sense of the shift is such that the IP beam must
be directed to us from the front of the star, and the MP's second
component is an emission beam from behind the star (Fig. \ref{fig2}).
To take into account retardation and aberration (but neglecting
gravitational bending and refraction) in a bidirectional model we may
use Eq. \ref{Eqtwopoleshift} with the differential height $\Delta h$
now replaced by $2h_{em}$, where $h_{em}$ is the distance of the
source from the neutron star 
(\citealt{drh04,dzg05}).
\begin{equation}
\label{Eqbidirectionalshift}
\Delta \phi = \frac{8\pi h_\mathrm{em}}{cP},
\end{equation}
Setting $\Delta \phi = \degrees{4}$, we obtain an emission height of
250 km for the source of the 'I/M' source, roughly consistent with
our estimate of the beam emission height in
Sect. \ref{twopolemodelfit}.  Such an emission height is not
incompatible with he wide range of estimates that observers and
theorists have made for pulsar emission heights over the years,
although it tends to be at the lower end.

We now need to explain the origin of the leading component of the
MP, and this is best done with reference to the fiducial
points marked on Fig. \ref{fig2} (upper diagram). These indicate the
 pulse longitudes at which the magnetic poles on the surface
of the star are directed to the observer. The peak of the leading
component  of the MP appears about 
\degrees{7} in advance of the modulated component of the MP,
which means that the unmodulated component 'U' must point towards us
about \degrees{5} of longitude ahead of the fiducial point on the
star,
giving it an altitude of some 600 km (Eq. \ref{Eqtwopoleshift} with
$\Delta h_\mathrm{em}$ replaced by $h_\mathrm{em}$). The emission beam
would have a radius of $\rho=\degrees{18}$ at this altitude if it is
confined by the last closed fieldlines (Eq. \ref{emissionheight}),
which corresponds to an expected component width of $W=\degrees{32}$
(Eq. \ref{cos_rho}). The observed width of the unmodulated component
is clearly narrower (Fig. \ref{polprofile}), showing that the model
requires the beam corresponding to this component to be considerably
narrower than the cone defined by the last closed fieldlines. 
This may not be an insuperable problem, since there is evidence that 
some pulsars indeed have beams which lie well within the last open 
fieldlines (\citealt{gg01}, \citealt{gg03}, \citealt{mr02a}).

 However attractive the bidirectional model may appear,
significant objections remain. Firstly, the model is asymmetric -- why
don't we see modulated emission from the second pole? Secondly, and
perhaps more seriously, the PA-swing from the leading to
trailing component of the MP is very smooth 
(Fig. \ref{polprofile}). If the two components originate from
opposite sides of the star,  aberrational and time-delay
effects would shift their  PA-swings out of phase
(\citealt{dfs+05}, \citealt{bcw91}). Possibly, the PA of the
unmodulated emission dominates the polarisation picture for
the MP, although this cannot  explain the IP emission, which
is highly modulated and is almost $100\%$ linearly polarised.
Thirdly, this model is purely geometric in nature and needs
to explain how coherent radiation of the same frequency is emitted at
roughly the same height. Upflow and downflow would need to be
simultaneously present, perhaps resulting from modulated pair-creation
in the magnetosphere.

\subsection{Comparisons with other interpulse pulsars}

\subsubsection{PSR B1822--09}

PSR B1822--09 offers an immediate comparison in many ways. Like PSR
B1702--19, this pulsar has a double-peak  MP and an
 IP approximately \degrees{180} away. Furthermore, its MP's
trailing component and IP are both periodically modulated with the
same $P_3$ $(\approx40P)$ (\citealt{fmw81}). It is not known whether
these components are closely phase-locked, as in PSR B1702--19, but if
this were to prove so, then many of our above arguments in favour of
an annular emission beam and a possible common source for the
modulated components might apply to this pulsar also.

It is possible to construct a bidirectional model for PSR B1822--09
along the same lines as PSR B1702--19 (Fig. \ref{fig2}). However,
there are significant differences in the component separations
(\citealt{gjk+94}), and this gives the model a different
structure. Firstly, in PSR B1822--09 the IP follows both the
components of the MP by more than \degrees{180}. When we apply the
logic of PSR B1702--19's model to PSR B1822--09 and insist that the
two modulated components are emitted from the same vector, 
then the IP must be the beam directed toward the star.
Secondly, the unmodulated leading component would point
towards the observer ahead of the common emission region for the MP
and IP (see Fig. \ref{fig2}).

The period of PSR B1822--09 is 0.77 seconds, more than twice that of
PSR B1702--19, and the derived height for the I/M source is about 1100
km (Eq. \ref{Eqbidirectionalshift}) and the leading component 6000 km
(Eq. \ref{Eqtwopoleshift} with $\Delta h_\mathrm{em}$ replaced by
$h_\mathrm{em}$). Note from Fig. \ref{fig2} that although the
arrangement of the components for this pulsar is somewhat different --
in that we seem to only see the emission from one side of the star --
the U component is again considerably farther from the star than the
I/M region and must therefore be narrower about the magnetic axis than
the periodically modulated region.

The model described above is somewhat different then the
\cite{dzg05} model, because they interpret the mode-changes of PSR
B1822--09 as reversals of the emission beam. PSR B1822--09 either
shows a strong leading MP component, or a strong IP and therefore they
associate the IP with the leading 'U' component of the MP, rather than
with the 'M' component.  If refraction and gravitational bending can
be neglected, the 'I/U' should be located about 3500 km above the
surface of the star.

\subsubsection{PSR B1055--52}

Quite apart from having an IP, this interesting pulsar has many
similarities to PSR B1702--19. It has the same inferred surface field
strength ($1.1\times10^{12}$ Gauss) and a period and timing age of
just half that of PSR B1702--19 (500 kyr and 0.20
seconds). Furthermore, the form of its MP, like PSR B1702--19,
consists of a strong leading component and an overlapping double
second component. \cite{big90a} even reports the detection of a weak
modulation feature of $P_3=10P$ in the second component of the MP,
despite his single pulse data having a weak $S/N$.  PSR B1055--52 is
also known to be a powerful source of X-rays (it is one of the three
Musketeers; e.g. \citealt{ch83},
\citealt{dcm+05}), which could ultimately lead to a better
understanding of its geometry.

As in PSR B1822--09, and possibly PSR B1702--19, the leading component
of PSR B1055--52's MP seems to be implicated in a mode-change (see
Fig.  2 of \citealt{big90a}). The pulsar's weak $S/N$ hitherto
prevented a more detailed single-pulse analysis, so the exact
implications of the mode-change for the other components and intensity
modulations are unknown.

If we assume that the modulated component of the MP and IP share a
common source, the model for PSR B1702--19 might apply to this pulsar
also. The structure and relative positioning of the MP and IP follow
PSR B1702--19 in having less than \degrees{180} between them, so a
bidirectional model based on their separations results in the same
relative pattern for their components (Fig. \ref{fig2}). The 'M'
component is placed at the location of the minimum in the middle of the
'conal' trailing component of the MP and the 'I' component is taken to
be the minimum in the middle of the IP component (see Fig. 1 of
\citealt{big90a}).  These component separations would imply an
emission height of 1300 km for the bidirectional vector and an
emission height of only 150 km for the unmodulated source.
It should be noted that the widths of the MP and IP of PSR B1055--52
are considerably larger than that of PSRs B1702--19 and B1822--09, and
therefore the precise pulse longitudes of the 'U', 'M' and 'I' peaks
are less clearly defined.

\section{\label{SctSummary}Summary}

We have investigated the properties of PSR B1702--19. The pulsar is
unusual in having both a main pulse (MP) and an interpulse (IP). Here
we give a summary of our results.

Both the MP and the IP are modulated with a period of
$P_3=10.4\pm0.3P$, comprising an intensity modulation with only weak
phase drift. There is a phase difference close to $0.5P$ between the
modulation of the trailing component in the MP and that in the IP,
demonstrating that these two regions of the pulsar are intrinsically
exactly in phase. This phase lock is held over many modulation cycles,
despite small secular changes in the periodicity. There is no evidence
from observations over widely separated epochs that the modulations
ever fall out of phase. Even when integrated over the emission cycle,
the intensities of the IP and the highly-modulated trailing component
of the MP are correlated.

It is difficult to explain the phase synchrony with ``carousel'' or
``patchy'' beams. It would have to be a coincidence that our traverses
of the MP and IP beams intersect regions which are exactly in
phase. This would suggest a beam structure with a modulated uniform
cone of emission, contrary to the expectations of much current theory.

The polarisation data implies that the magnetic axis of PSR B1702--19
rotates almost perpendicularly to its rotation axis, which means that
models with a single or double emission cones close to alignment can
be eliminated.

If the modulated emission in the MP and IP come from two different
magnetic poles, then to keep them in the observed synchrony only
models with virtually instantaneous communication between the poles
are possible. These may involve processes which have attracted little
theoretical attention in the past, such as non-radial oscillations of
the star.

If the modulated emission is assumed to be oppositely directed on the
same fieldline on the same side of the star, it seems possible to
construct a model which accounts for the slight deviations of these
components from \degrees{180} separation in terms of aberration and
retardation. It would mean the IP emission is directed upwards and the
modulated component of the MP directed downwards and observed from
behind the star. The unmodulated emission from the MP would be
directed upwards on the opposite magnetic pole to the sources of the
modulated emission.  However, such a model requires the
emission of the MP to originate from two different poles, which is
difficult to reconcile with the observed smooth PA-swing. Whether the
bidirectional or the two-pole model turns out to be correct, the
answer will have important implications for emission theories.  

There is a striking similarity between PSRs B1702--19 and
B1822--09, a pulsar which also shows periodic modulations at the MP
and IP. If a phase lock operates between them as strictly as in PSR
B1702--19, then the same problems of interpole communication will
apply.

\begin{acknowledgements}
We would like to express our gratitude to J.~H. Seiradakis and
A. Jessner, who kindly made available archival Effelsberg data and
helped us with the data reduction.  GAEW thanks the Netherlands
Foundation for Scientific Research (NWO) and the Anton Pannekoek
Institute, Amsterdam, for their kind hospitality and the University of
Sussex for a Visiting Fellowship.  The Westerbork Synthesis Radio
Telescope is operated by the ASTRON (Netherlands Foundation for
Research in Astronomy) with support from NWO.
\end{acknowledgements}


\end{document}